\documentclass[preprint,aps]{revtex4}

\usepackage{graphicx}
\usepackage{dcolumn}
\usepackage{bm}

\usepackage{epsf}
\usepackage{anysize}
\usepackage{graphicx}
\marginsize{1 in}{1 in}{1in}{1.40in}

\usepackage{graphicx}
\usepackage{dcolumn}
\usepackage{bm}
\usepackage{amsmath}
\usepackage{amssymb}
\usepackage{mathrsfs}
\usepackage[latin1]{inputenc}
\usepackage{amsmath}

\makeatletter
  \newcommand\figcaption{\def\@captype{figure}\caption}
  \newcommand\tabcaption{\def\@captype{table}\caption}
\makeatother

\newcommand{\dd}{{\rm d}}

\newcommand{\J}{\mathcal{J}}
\newcommand{\lmd}{\lambda }
\newcommand{\om}{\omega}

\newcommand{\D}{\mathcal{D}}
\newcommand{\G}{{\rm{Stab}}}
\newcommand{\U}{\mathcal{U}}
\newcommand{\n}{\mathcal{N}}
\newcommand{\SUM}{{\rm SUM}}

\newcommand{\h}{\mathcal{H}}

\newcommand{\R}{{\mathbb{R}}}
\newcommand{\tr}{{\rm Tr}}

\newcommand{\Sp}{{\rm Sp}}
\newcommand{\OSp}{{\rm OSp}}
\newcommand{\spl}{{\rm sp}}

\newtheorem{property}{Propoerty}
\newtheorem{theorem}{Theorem}
\begin {document}


\title{Critical Topology for Optimization on the Symplectic Group}

\author{Rebing Wu}
\altaffiliation{Department of Chemistry, Princeton University,
Princeton, New Jersey 08544, USA}


\author{Raj Chakrabarti}
\altaffiliation{Department of Chemistry, Princeton University,
Princeton, New Jersey 08544, USA}

\author{Herschel Rabitz}
\altaffiliation{Department of Chemistry, Princeton University,
Princeton, New Jersey 08544, USA}

\begin{abstract}
Optimization problems over compact Lie groups have been
extensively studied due to their broad applications in linear
programming and optimal control. This paper analyzes least square
problems over a noncompact Lie group, the symplectic group
$\Sp(2N,\R)$, which can be used to assess the optimality of
control over dynamical transformations in classical mechanics and
quantum optics. The critical topology for minimizing the Frobenius
distance from a target symplectic transformation is solved. It is
shown that the critical points include a unique local minimum and
a number of saddle points. The topology is more complicated than
those of previously studied problems on compact Lie groups such as
the orthogonal and unitary groups because the incompatibility of
the Frobenius norm with the pseudo-Riemannian structure on the
symplectic group brings significant nonlinearity to the problem.
Nonetheless, the lack of traps guarantees the global convergence
of local optimization algorithms.
\end{abstract}

\maketitle

\section{Introduction}

The topology of solution sets to problems in the calculus of
variations is the subject of considerable interest in mathematical
physics and optimization theory \cite{HelMor1994}. It is of
particular importance in theory of optimal control, where this
topology can affect the efficiency of the search for effective
control Hamiltonians \cite{RabMik2004}. Whereas in general it is
very difficult to characterize these features for arbitrary
functionals, when the objective or Lagrangian functional is
defined on a Lie group, it is often possible to apply techniques
from the theory of Lie groups and differential geometry to
simplify the extraction of critical topology.


The topology of the critical submanifolds of classical Lie groups
was originally studied by Frankel \cite{Frankel1965}, who
characterized the number of critical points and associated Morse
indices of the trace function on compact classical Lie groups
$\U(N)$, $O(N)$, and $\Sp(N)$. Dynnikov and Vesselov
\cite{Dynnikov1997} subsequently identified these functions as
perfect Morse-Bott functions and showed that they afford a cell
decomposition of the associated groups. Recently, the equivalence
of the trace function to that of a least-square matrix function
for the distance between a real and target transformation led to
the application of these results to optimization and control
theory. Brockett \cite{Brockett1989} showed that a wide range of
combinatorial optimization problems arising in linear programming
can be framed as matrix least squares optimizations on compact Lie
groups. In \cite{RabMik2005}, the critical topology of the trace
function on $\U(N)$ was analyzed in light of its connection to the
optimal control problem of implementing a quantum logic gate over
discrete variables with maximal fidelity.



A unifying feature of these problems is the fact that the domain
of the objective functional, being a compact Lie group, can always
be endowed with the structure of a differential manifold with a
bi-invariant Riemannian metric. In this paper, we attempt to
extend such studies to the investigation of critical submanifolds
of least squares objective functions on noncompact Lie groups
\cite{Mahony2002}, in particular the symplectic group
$\Sp(2N,\R)$. Although the geometry and topology of symplectic
manifolds, and functions defined on those manifolds, have been the
subject of extensive investigations in mathematical physics,
functions defined on the symplectic group itself have received far
less attention.

Specifically, we are concerned with the least-square distance
function on the space of symplectic matrices,
\begin{equation}\label{symplectic landscape}
\J(S)=\|S-W\|^2=\tr(S-W)^T(S-W), \quad S\in \Sp(2N,\R).
\end{equation}
This cost function has recently been shown to have fundamental
applications in the assessment of the fidelity of dynamical gates
in quantum analog computation when implemented through optimal
control theory \cite{WuRaj2007}, where $W$ represents the target
quantum gate to be realized. Another potential important
motivation for studying this problem comes from the control of
beam systems in particle accelerators \cite{DraNer1988}. As shown
before, the cost function (\ref{symplectic landscape}) on compact
Lie groups (e.g., $\U(N)$, $O(N)$) is equivalent to a linear trace
function. However, this no longer holds on the symplectic group,
because the corresponding Riemannian metric is not bi-invariant
under symplectic transformations. This feature is caused by
incompatibility of the Frobenius norm defined in (\ref{symplectic
landscape}) with the geometric structure of the symplectic group,
and greatly complexifies the critical topology, as we will show
below. On the other hand, this group can be treated as a
pseudo-Riemannian manifold with a bi-invariant pseudo-Riemannian
metric. Although it is possible to introduce an objective function
that is compatible with this pseudo-Riemannian metric
\cite{CarSil2006}, such function is not positive definite and
cannot be interpreted as a distance function. However, the
corresponding critical topology is equivalent to that of a linear
trace function on $\Sp(2N,\R)$, as well as those on $\U(N)$ and
$O(N)$. In contrast to the objective functionals (\ref{symplectic
landscape}), such compatibility leads to a simple critical
topology as the effects of the pseudo-Riemannian geometry of
noncompact Lie groups.

Existing works on control of classical mechanical systems
generally do not require direct control of the system propagators
except some special cases (e.g., robotic motion planning on the
Euclidean group $SE(3)$ \cite{HanPark2001}). It is usually
sufficient to attain the control over state vector with fewer
degrees of freedom.
However, since any given state vector in phase space is associated
with an infinite number of symplectic matrices that propagate the
initial state of the system to the desired final state, it is
generally impossible to predict which of these symplectic matrices
will be reached by the time-dependent control obtained through the
optimization procedure
As such, the efficiency of control optimization will be highly
system-dependent, with optimization algorithms traversing longer
trajectories in the symplectic group for certain classes of
Hamiltonians. In contrast, if the control problem is cast in terms
of symplectic propagator optimization, it is possible to choose
the shortest path in the symplectic group from the initial
condition to the target \footnote{In this approach, distance is
measured in terms of the length of the geodesic joining these two
matrices in the group.}. As such, gradient control algorithms
based on propagator optimization may outperform those based on
state vector optimization. Optimization algorithms of this type
are currently the
subject of intense study in the context of quantum control \cite{RajWu2007, KhaBro2001, KhaBro2002, Glaser1998}. 

In a study of optimization algorithms on noncompact Lie groups,
Mahony indicated that (local) quadratic convergence can still be
achieved by using the Newton method adapted for the curved
manifold under local coordinates of the first kind
\cite{Mahony2002}, which results in no essential differences
compared to algorithms on compact Lie groups. However, the global
topology of the optima and suboptima may play a fundamental role
in the overall efficiency of the optimizations, and will be the
major concern of our studies here. This paper is organized as
follows. Section II summarizes the definition and properties of
symplectic groups. Section III derives the canonical form of
landscape critical points. Section IV analyzes the Hessian
quadratic form for each of these critical points. Section V
studies the constrained landscape over the compact subgroup.
Section VI provides an illustrative example. Finally, Section VII
draws the conclusion.
\section{Preliminaries on the symplectic group}
In classical mechanics, a transformation for a system described by
$N$ pairs of coordinate and momentum variables
$z_{\alpha,\beta}=(x^1_{\alpha,\beta},\cdots,x^N_{\alpha,\beta};p^1_{\alpha,\beta},\cdots,p^N_{\alpha,\beta})$
is called symplectic if it preserves the (skew-symmetric)
symplectic form
$$\omega(z_\alpha,z_\beta)=\sum_{i=1}^N(x_\alpha^ip_\beta^i-x_\beta^ip_\alpha^i).$$With this
coordinate system, a symplectic transformation can be represented
by a $2N\times 2N$ dimensional real matrix that satisfies
$S^TJS=J$, where $S^T$ is the transpose of $S$ and
$$J=\left(%
\begin{array}{cc}
   & I_N \\
  -I_N &  \\
\end{array}%
\right).
$$The set of symplectic matrices forms a noncompact Lie group $\Sp(2N,\R)$,
and its Lie algebra $\spl(2N,\R)=\{JB|B^T=B,~~B\in \R^{2N\times
2N}\},$ from which it is easy to see that the dimension of
$\Sp(2N,\R)$ is $N(2N+1)$. As a linear vector space, $\spl(2N,\R)$
can be decomposed into two mutually orthogonal subspaces
$\spl(2N,\R)=\mathcal{L}_1\oplus\mathcal{L}_2$, where
\begin{eqnarray*}
  \mathcal{L}_1 &=& \{JB\in
\spl(2N,\R)|JB=-BJ\}, \\
  \mathcal{L}_2 &=& \{JB\in \spl(2N,\R)|JB=BJ\}.
\end{eqnarray*}
The subspace $\mathcal{L}_2$ is a Lie subalgebra of $\spl(2N,\R)$.
It generates the orthogonal symplectic group $\OSp(2N,\R)$ as the
maximal compact Lie subgroup of $\Sp(2N,\R)$, as it is the
intersection of the symplectic group $\Sp(2N,\R)$ with the
orthogonal group $O(2N)$. There is an interesting isomorphism
between $\OSp(2N,\R)$ and the unitary group $\U(N)$ via the
following mapping:
\begin{equation}\label{U-S map}
X-iY\rightarrow \left(%
\begin{array}{cc}
  X& Y \\
  -Y & X \\
\end{array}%
\right),\quad X-iY\in \U(N).\end{equation}

Here we briefly summarize some properties of symplectic matrices and
symplectic groups that will be used in the following analysis.
Readers of interests are referred to \cite{DraNer1988} for more
details.
\begin{property}
As the analog of the property $O^{-1}=O^T$ for any orthogonal
matrix $O$, $S^{-1}=J^TS^TJ$ for any symplectic matrix $S$.
\end{property}

\begin{property}
The eigenvalues of a symplectic matrix always appear in reciprocal
pairs, i.e., if $\om$ is an eigenvalue of $S$, then so is
$\om^{-1}$, and they have identical degeneracy degrees.
\end{property}

\begin{property}
There always exists a symplectic singular value decomposition
(SVD) $S=UDV$, where $U$ and $V$ are orthogonal symplectic
matrices. $D$ is a diagonal symplectic matrix whose diagonal
elements are the singular values of $S$.
\end{property}

\begin{property}
Denote by $\G(D)=\{R\in\OSp(2N,\R):R^TDR=D\}$ the stabilizer of
the diagonal symplectic matrix $D$ in the group $\OSp(2N,\R)$. The
stabilizer of a diagonal symplectic matrix
$$D=diag\{a_0I_{n_0},a_1I_{n_1},\cdots,a_rI_{n_r};a_0I_{n_0},a_1^{-1}I_{n_1},\cdots,a_r^{-1}I_{n_r}\},$$
where $1=a_0<a_1<a_2\cdots<a_r$, is the direct product of
subgroups $\G(D)=\OSp(2n_0,\R)\times O(n_1)\times\cdots\times
O(n_r).$
\end{property}

The above features can be demonstrated by the example of the
2-qunit SUM gate in continuous quantum computation
\cite{WuRaj2007,BarSan2002, GotKit2001}, which acts on the
quadratic vector $(q_1,q_2;p_1,p_2)$ as follows
$$\SUM:\quad q_1 \rightarrow q_1,\quad q_2 \rightarrow q_1 + q_2,\quad p_1 \rightarrow  p_1 - p_2,\quad p_2 \rightarrow p_2.$$
The matrix form of the SUM gate is
\begin{equation}\label{sum}
\SUM =\left(%
 \begin{array}{cccc}
  1 & 0 & 0 & 0 \\
  1 & 1 & 0 & 0 \\
  0 & 0 & 1 & -1\\
  0 & 0 & 0 & 1 \\
\end{array}%
\right),
\end{equation}
whose singular value decomposition $\SUM=UEV$ can be found to be
\begin{equation}\nonumber
U= \left(%
\begin{array}{cccc}
         -\xi &  -\eta &  0 &      0 \\
        -\eta &  \xi &  0 &      0 \\
        0 &        0 &  -\xi &-\eta  \\
        0 &        0 &  -\eta &   \xi \\
\end{array}%
\right),~~
E= \left(%
\begin{array}{cccc}
\om &&& \\
&\om^{-1}&& \\
&&\om^{-1}& \\
&&&\om \\
\end{array}%
\right),~~ V=\left(%
\begin{array}{cccc}
 -\eta &   -\xi & 0 &        0 \\
\xi &  \eta     &    0 &        0      \\
        0 &        0 & -\eta &  -\xi     \\
         0 &        0 &  \xi &  \eta \\
\end{array}%
\right),
\end{equation}
where $$\xi=\sqrt\frac{5-\sqrt{5}}{10},\quad
\eta=\sqrt\frac{5+\sqrt{5}}{10},\quad \om=\frac{\sqrt{5}+1}{2}.$$
Since there is a two-fold degeneracy of the singular value $\om$,
the stabilizer of $E$ is isomorphic to the $O(2)$ group.

\section{Canonical Form of the critical submanifolds}
Any candidate solution to the optimization problem
(\ref{symplectic landscape})must be one of its critical points,
defined as a $S^*\in\Sp(2N,\R)$ such that the gradient $\nabla
J(S^*)$ of the cost function vanishes. Since there generally exist
multiple-solutions for the critical points, a complete
understanding of the critical topology is of essential importance
to assess the complexity of searching for the global optimal
solution.

The basic idea to determine the set of critical solutions is, at
an arbitrary fixed point $S\in\Sp(2N,\R)$, to perturb the cost
function along an arbitrary direction in the tangent space
(isomorphic to $\spl(2N,\R)$) and find those points where the
directional derivation vanishes along all directions. For example,
taking the parametrization $Se^{tJY}$ with $Y^T=Y$ (here $JY$
represents the local Cartesian coordinates in the tangent space at
$S$), the critical condition can be obtained by forcing the
derivative along $JY$ to be zero at $t=0$ for arbitrary $Y$, i.e.,
$$\frac{\dd }{\dd t}\J(Se^{tJY})\Big|_{t=0}= \tr[JY(S^TS-W^TS)]=0, \quad \forall~ Y^T=Y,$$which implies that the matrix $(S^TS-W^TS)J$ has to be skew-symmetric, i.e.,
\begin{equation}\label{b} (S^TS-W^TS)J=J(S^TS-S^TW).
\end{equation}
Left multiplying a constant matrix $J^T$ and applying the property
$J^TS^TJ=S^{-1}$, we get a simpler form:
\begin{equation}\label{critical condition}
S^TS-(S^TS)^{-1}=S^TW-(S^TW)^{-1}.
\end{equation}

Although (\ref{critical condition}) is a nonlinear (fourth-order
in $S$) equation, its highly symmetric form makes it still
solvable. Let $S=U_1DV_1$ be a symplectic SVD of $S$, and
$W=U_0E_dV_0$ be that of $W$. Substituting these SVDs into the
equation (\ref{critical condition}), we can simplify the condition
as
$$D^2-D^{-2}=DE-(DE)^{-1},$$
where $E=UE_dV$ with $U=U_1^TU_0$ and $V=V_0V_1^T$. Followed by a
commutation with $DE$ on both sides, this equation is further
transformed as $[DE,D^2-D^{-2}]=0$, or equivalently,
$[E,D^2-D^{-2}]=0$. We will show that this relation implies
$[E,D]=0$. Let $d_r>\cdots>d_1\geq1\geq d_1^{-1}>\cdots<d_r^{-1}$
be the distinct eigenvalues of $D$, where the degeneracy of $d_i$
(or $d_i^{-1}$) is $M_i$, $i=1,\cdots,r$; then $D$ can be
decomposed into diagonal blocks $D_i=d_iI_{M_i}$, $i=1,\cdots,r$,
and their inverses. The commutativity of $E$ and $D^2-D^{-2}$
implies that $E$ is simultaneously block-diagonal with
$D^2-D^{-2}$, corresponding to sub-blocks
$(d_i^2-d_i^{-2})I_{M_i}$. Since $D$ is positive definite, two
distinct eigenvalues of $D$ must correspond to two distinct
eigenvalues of $D^2-D^{-2}$, and vice versa. So $D$ shares the
same eigenspace decomposition with $D^2-D^{-2}$, as well as that
of $E$. Hence $E$ commutes with $D$.

Therefore, for each $D_i$ in $D$, there corresponds a diagonal
block $E_i$ of $E$ . Let $E_i=U_iE_{0i}V_i$ be a SVD of $E_i$,
then the following block-diagonal symplectic orthogonal matrices
\begin{eqnarray*}
U_d   &=& diag\{U_1,\cdots,U_r;U_1^T,\cdots,U_r^T\}, \\
V_d   &=& diag\{V_1,\cdots,V_r;V_1^T,\cdots,V_r^T\}, \\
E_d   &=&
diag\{E_{01},\cdots,E_{0r};E_{01}^{-1},\cdots,E_{0r}^{-1}\}
\end{eqnarray*} define a SVD $E=U_dE_dV_d$ of $E$.
Hence, for any general SVD $E=UE_dV$, the following relationship
$$(U_d^TU)E_d(VV_d^T)=E_d,$$ implies the existence of a symplectic orthogonal matrix $R\in
\G(E_d)$ such that $U=U_dR$ and $V=R^TV_d$.

Going back to the original symplectic matrix $S^*$, we then have a
uniform expression for the critical solutions:
$$S^*=U_1DV_1=U_0U^TDV^TV_0=U_0R^T(U_d^TDV_d^T)RV_0.$$
Notice that $U_d^T$ commutes with $D$, by which we can denote
$L=U_d^TV_d^T$, and then simplify the representation of the
critical points as folows:
\begin{equation}\label{critical points} S^*=U_0R^TPRV_0,\quad
P=DL,~R\in\G(E_d).
\end{equation}
This canonical form shows that the critical manifold consists of
orbits of admissible matrices $P$ under the action of $\G(E_d)$,
which can be represented by the quotient set
$\mathcal{M}={\G(E_d)}/{\G(P)}$. It is then sufficient to
characterize the set of critical points by specify all possible
values of the characteristic matrix $P$ that involves the singular
values $d_i$ and their corresponding orthogonal matrix blocks
$L_i=U_i^TV_i^T$.

For simplicity, owing to the reciprocal properties of singular
values of symplectic matrices, we only need to analyze the
singular values that are no less than 1. Restricting the matrix
equation (\ref{critical condition}) on the eigenspace of each
$d_i\geq 1$, and substituting the canonical form into
(\ref{critical condition}), we have
\begin{equation}\label{L-equation}
(d_i^2-d_i^{-2})I_{M_i}=d_iE_{0i}L_i-d_i^{-1}L_i^TE_{0i}^{-1}.
\end{equation}  Let $e_{1},\cdots,e_{M_i}$ be
the singular values of $E_{0i}$. This equation can be decomposed
as
\begin{equation}\label{L2} d_ie_\alpha L
_{i,\alpha\beta}-d_i^{-1}e_\beta^{-1}L_{i,\beta\alpha}=(d_i^2-d_i^{-2})\delta_{\alpha\beta},
\quad  \alpha,\beta= 1,\cdots,M_i,
\end{equation}where $L
_{i,\alpha\beta}$ is the $\alpha\beta$-th matrix element of $L_i$.
Using equation (\ref{L2}), we classify the sub-blocks in $P$ into
the following different types.

Firstly, suppose that the orthogonal matrix $L_i$ is diagonal,
then each element $L_{i,\alpha\alpha}$ has to be unimodular. When
$L_{i,\alpha\alpha}=1$, we get
$$(d_i-e_\gamma)(d_i^3+e_\gamma^{-1})=0, $$
whose only admissible positive root is $d_i=e_\gamma$, where
$e_\gamma\geq1$. The case $L_{i,\alpha\alpha}=-1$ corresponds to
$$(d_i+e_\gamma)(d_i^3-e_\gamma^{-1})=0, $$
of which the only admissible positive root is
$d_i=e_\gamma^{-1/3}$ where $e_\gamma<1$.

In such cases, each block $E_i$ allows for only one singular value
so that all eigenvalues of $D_i$ are identical. Corresponding the
case $L_{i}=I_{M_i}$, the block $D_i$ is called type I; for the
case $L_{i}=-I_{M_i}$, the block $D_i$ is called type II.

For the more general case that $L_i$ is not diagonal, any pair of
nonzero off-diagonal matrix elements must satisfy
\begin{equation}\label{nonzero diagonal}
\left(%
\begin{array}{cc}
  d_ie_\alpha  & -d_i^{-1}e_\beta^{-1} \\
  -d_i^{-1}e_\alpha^{-1} & d_ie_\beta \\
\end{array}%
\right)\left(%
\begin{array}{c}
  L_{i,\alpha\beta} \\
  L_{i,\beta\alpha} \\
\end{array}%
\right)=\left(%
\begin{array}{c}
 0 \\
 0 \\
\end{array}%
\right),\quad \alpha\neq \beta,\end{equation} in which the
determinant of the coefficient matrix has to vanish, and this
solves the eigenvalue $d_i$, i.e.,
\begin{equation}\label{off-diagonal }
d_i^2e_\alpha e_\beta-(d_i^2e_\alpha
e_\beta)^{-1}=0~\Longleftrightarrow ~d_i=(e_\alpha
e_\beta)^{-1/2}.
\end{equation}
Consequently, substituting (\ref{off-diagonal }) back into the
equation (\ref{nonzero diagonal}), we find that the resulting
nonzero off-diagonal matrix element
$L_{i,\alpha\beta}=L_{i,\beta\alpha}$, i.e., the matrix $L_i$ must
be symmetric. Obviously, in such cases each minimal block $D_{i}$
allows for exactly two distinct singular values of $E_i$
(otherwise $D_i$ will have non-unique singular values), and their
repeating number are both $k_i=M_i/2$.

Without loss of generality, we assume that $e_\alpha\geq e_\beta$
and $d_i=(e_\alpha e_\beta)^{-1/2}\geq1$. Then the use of
(\ref{off-diagonal }) solves the corresponding diagonal elements
from (\ref{L2}) as follows
$$L_{i,\alpha\alpha}=-L_{i,\beta\beta}=\frac{d_i^2-d_i^{-2}}{d_ie_\alpha -d_i^{-1}e_\alpha^{-1}}
=\frac{(e_\alpha e_\beta)^{-1}-e_\alpha
e_\beta}{(e_\alpha/e_\beta)^{1/2} -(e_\beta/e_\alpha)^{-1/2}}\geq
0.
$$
As $L_i$ is orthogonal, each of its matrix elements must satisfy
$0<L_{i,\alpha\beta}\leq 1$, which set additional constraints on
the admissible pairs $e_\alpha$ and $e_\beta$ that generate $d_i$:
\begin{equation}\label{di condition}
e_\beta\leq1\leq e_\alpha,\quad e_\beta^{-1/3}\leq e_\alpha\leq
e_\beta^{-1}.
\end{equation}

Under such conditions, the corresponding block $D_i$ is called a
type III block. Suppose that $E_i=diag\{
  e_\alpha I_{k_i},e_\beta I_{k_i}\}$, the corresponding matrix $L_i$ must be in the
following form:
$$L_i=\left(%
\begin{array}{cr}
  \cos x_i~ I_{k_i} &  \sin x_i~ O^T_{k_i}\\
  \sin x_i ~O_{k_i}& - \cos x_i~ I_{k_i}\\
\end{array}%
\right),$$ where $O$ is some orthogonal matrix and the angle
$x_i=\arccos \frac{(e_\alpha e_\beta)^{-1}-e_\alpha
e_\beta}{(e_\alpha/e_\beta)^{\frac{1}{2}}
-(e_\beta/e_\alpha)^{-\frac{1}{2}}}\in[0,\frac{\pi}{2}].$ Let
$T=diag\{O,O\}$, then $P^TL_iP$ is in the following canonical form
$$L_i=\left(%
\begin{array}{cr}
  \cos x_i I_{k_i}&  \sin x_iI_{k_i}\\
  \sin x_i I_{k_i}& - \cos x_iI_{k_i}\\
\end{array}%
\right).$$Since the transformation matrix $T$ is in the stabilizer
$\G(E_d)$, so $L_i$ can be always represented by the above
standard form.

In conclusion, suppose that $W$ has $s$ greater-than-1 singular
values $1<\om_1< \om_2< \cdots< \om_s$ with degeneracy degrees
$n_1,\cdots,n_s$. From the above analyses, each given singular
value $\om_\alpha>1$ can be used to produce a singular value $d_i$
in the canonical form $P$ through the following three ways:

\begin{enumerate}
    \item $d_i=\om_\alpha$ with multiplicity $m'_\alpha$ and a corresponding
matrix block $P'_\alpha=\om_\alpha I_{m'_\alpha}$ in $P$;
    \item
$d_i=\om_\alpha^{-\frac{1}{3}}<1$ with multiplicity $m''_\alpha$
and a corresponding matrix block
$P''_\alpha=-\om_\alpha^{-\frac{1}{3}} I_{m''_\alpha}$ in $P$;
  \item
$d_i=(\om_\alpha/\om_\beta)^{-1/2}$ with multiplicity
$2m'''_{\alpha\beta}$, where $\om_\alpha\preceq \om_\beta$
(defined as that $\om_\beta^{1/3}\leq \om_\alpha \leq \om_\beta$),
which are from $m'''_{\alpha\beta}$ eigenvalues $\om_\alpha$ and
$m'''_{\alpha\beta}$ eigenvalues $\om_\beta$. The corresponding
block is
$$P'''_{\alpha\beta}=-\left(\frac{\om_\alpha}{\om_\beta}\right)^{-\frac{1}{2}}\left(%
\begin{array}{cr}
  \cos x_{\alpha\beta}I_{m'''_{\alpha\beta}} &  \sin x_{\alpha\beta}I_{m'''_{\alpha\beta}}\\
  \sin x_{\alpha\beta}I_{m'''_{\alpha\beta}}& - \cos x_{\alpha\beta}I_{m'''_{\alpha\beta}}\\
\end{array}%
\right),$$ where $x_{\alpha\beta}=\arccos\frac{(\om_\alpha/
\om_\beta)^{-1}-\om_\alpha/\om_\beta}{(\om_\alpha
\om_\beta)^{\frac{1}{2}} -( \om_\alpha
\om_\beta)^{-\frac{1}{2}}}.$
\end{enumerate}

The singular value $\om_0=1$ can be treated separately. Since the
condition (\ref{di condition}) can never be satisfied with
$\om_0=1$ and any other singular values of $E$, $\om_0=1$ is
always incapable of generating a type III singular value $d_i\neq
1$ via (\ref{off-diagonal }). Thus, they only contributes to type
I or II singular values $d_i=1$ of $D$. Let the degeneracy number
of $\om_0=1$, which must be even, be $2n_0$, and the number of
type I singular values is $m_0$, the possible characteristic
matrix block $P_0$ is
$$P_0=\left(%
\begin{array}{cc}
  I_{m_0} &  \\
   & -I_{n_0-m_0} \\
\end{array}%
\right),\quad m_0=0,\cdots,n_0.$$

Any group of admissible indices
$\{m_0;m'_\alpha,m'''_\alpha,m'''_{\alpha\beta};\alpha,\beta=1,\cdots,s\}$
labels an orbit of $\G(E_d)$, which equivalently labels a unique
critical submanifold of the set of critical points. The value of
the cost function at these critical submanifolds are:
\begin{eqnarray*}
  J(S^*)& = &8(n_0-m_0)^2+\sum_\mu
   m''_\mu(\om_\mu^{2}+\om_\mu^{-2}+3\om_\mu^{2/3}+3\om_\mu^{-2/3})\\
&&
+\sum_{\alpha\preceq\beta}m'''_{\alpha\beta}[(\om_\alpha+\om_\beta^{-1})^2+(\om_\alpha^{-1}+\om_\beta)^{2}].
\end{eqnarray*}

\section{Topology analysis of critical submanifolds}
This section will delve into more intrinsic topological details of
the critical manifolds, including (1) their connectedness,
determined by counting the number of separate submanifolds, (2)
their dimensions and local optimality status (i.e, local maximum,
minimum or saddle point), determined via Hessian analysis. Such
information provides a global picture of the distribution of
possible solutions and their influences on the actual search for
optimal solutions to the optimization problem.

The number of critical submanifolds can be enumerated by counting
all admissible combinations of indices
$\{m_0;m'_\alpha,m''_\alpha,m'''_{\alpha\beta};\alpha,\beta=1,\cdots,s\}$,
each of which corresponds to a unique characteristic matrix $P$,
and hence labels a critical submanifold as the orbit of $P$ under
the action of $\G(E_d)$ \footnote{Note that $\G(E_d)$ can be a
disconnected manifold because its subgroups $O(n_i)$ are not
connected. However, the orbit of $P$ under the actions of
different branches of $\G(E_d)$ coincide with each other. So the
orbit as a critical submanifold is still connected.}. This number
is dependent with the degenerate structure of the singular values
of the target transformation $W$. For example, the simplest case
is that $W$ is an orthogonal symplectic matrix, where $E_d=I_{2N}$
and $P$ has only $\pm1$ singular values. The total number of
critical submanifolds is $N+1$ corresponding to $m=0,1,\cdots,N$,
the repeating number of the $-1$ eigenvalues in $P_0$. When $W$
has a fully degenerate singular value $\om>1$, the admissible
characteristic matrices $P$ may contain either of the I-III types
of sub-blocks, and hence there are more critical submanifolds. Let
$m'''$ be the number of pairs that generate $2m'''$ type III
singular values $d=1$ in $D$, $m'$ for $d=\om$ and $m''$ for
$d=\om^{1/3}$. Since $m'+m''+2m'''=N$, counting such admissible
combinations gives the number of critical submanifolds as a
quadratic function of $N$:
\begin{equation}\label{enum}
  \n=\left\{\begin{array}{ll}
  (N+2)^2/2, & ~N~~{\rm even}; \\
(N+1)(N+3)/2, &~ N ~~{\rm odd}; \\
\end{array}\right.
\end{equation}

The number of critical submanifolds shoots up when the degeneracy
in $W$ is broken up. The extremal case is that $E_d$ is fully
non-degenerate and the singular values of $E_d$ are not far apart
from each other such that $\om_\alpha\preceq \om_\beta$ for any
$\alpha<\beta$, i.e., any two distinct singular values of $W$ are
allowed to produce a pair of type III singular values of $S^*$.
Let $m$ be the number of pairs of singular values of $W$ that
generate type III singular values of $S^*$; then there are
$N!/[(N-2m)!m!2^m]$ different choices. Moreover, for each fixed
$m$, the possibilities of using the remaining singular values to
generate I or II type singular values of $S^*$ is $2^{N-2m}$.
These set an upper bound
$$\mathcal{N}=\sum_{m=1}^{[N/2]}\frac{2^{N-3m}N!}{m!(N-2m)!},$$which is super-exponential in $N$, on the maximal
number of critical submanifolds in all cases.

The dimensions of the critical submanifolds are generally
difficult to calculate. However, are simple for those that
contains only type I and II blocks (i.e., $m'''_{\alpha\beta}=0$,
for all $\alpha,\beta$) via their geometrical expression
$\mathcal{M}=\G(E_d)/\G(P)$, i.e.,
$${\rm dim}\mathcal{M}= {\rm dim}\G(E_d)-{\rm dim}\G(P).$$
As stated in Section II, $\G(E_d)$ is the product of orthogonal
subgroups $O(n_\mu)$ and a symplectic orthogonal group
$\OSp(2n_0,\R)$ (for $\om_0=1$). The stabilizer of $P$ is a Lie
subgroup of $\G(E_d)$, which is the product of $O(m'_\mu)$ and
$O(m''_\mu)$ for type I and II singular values. Therefore, such
critical submanifolds can be represented as
$$\mathcal{M}=\frac{\OSp(2n_0,\R)}{\OSp(2m_0,\R)\times \OSp(2n_0-2m_0,\R)}\times \prod_{\mu=1}^s\frac{ O(n_\mu)}
{O(m'_{\mu})\times O(m''_\mu)},$$ and their dimensions can be
easily evaluated as
\begin{equation}
 \mathcal{D}  = 2m_0(n_0-m_0)+\sum_{\mu=1}^s{m'_\mu}{m''_\mu}.
\end{equation}

The optimality status of these critical submanifolds can be
acquired from analysis of the local geometric structure for each
of the critical submanifolds via their Hessian quadratic form
(HQF). The numbers of positive, negative and zero Hessian
eigenvalues determine the optimality status, i.e., a critical
point is a local minimum (maximum) if all the eigenvalues are
positive (negative), otherwise it is a saddle point. The HQF is
defined as the second-order term of $Y$ in the Taylor expansion of
the parametrization $Se^{tJY}$, which is dominant in the
neighborhood of $S^*$ while the first-order term vanishes at $S$.
It is not difficult to obtain that
\begin{eqnarray*}
  \h(Y) &=& \tr~[JY(S^TS-W^TS)JY+JYS^TS(JY)^T] \\
   &=& \tr~[JYV_0^TR^T(P^2-E_dP)RV_0JY+JYV_0^TR^TP^2RV_0(JY)^T]
\end{eqnarray*}
Notice that (1) $P^2=D^2$ because $L$ commutes with $D$ and $L$ is
symmetric orthogonal; (2) $JRV_0=RV_0J$, we may transform $Y$ into
$X=(RV_0)Y(RV_0)^T$ and rewrite the HQF as
\begin{equation}\label{HQF-X}
    \h(X)=\tr~[JX(D^2-DE_dL_d)JX+JXD^2(JX)^T].
\end{equation}Let ${\bf\rm x}$ be the vector of independent variables in $X$, then $\h(X)$ can be written as a quadratic
form ${\bf\rm x}^T \mathcal{Q} {\bf\rm x}$, where $\mathcal{Q}$ is
a symmetric $N(2N+1)\times N(2N+1)$ matrix. The Hessian
eigenvalues are defined as the eigenvalues of the matrix
$\mathcal{Q}$.

Let $D=diag\{\Theta,\Theta^{-1}\}$,
$DE_d=diag\{\Omega,\Omega^{-1}\}$ and $L=diag\{\Phi,\Phi\}$.
Dividing the symmetric matrix $X$ as
$$X=\left(%
\begin{array}{cc}
  A & C^T \\
  C & B \\
\end{array}%
\right),$$ where $A$ and $B$ are symmetric, we may rewrite the HQF
as the function of $A$, $B$ and $C$, i.e.,
$$\h(A,B,C)=\tr(A\Theta^2A-2A\Sigma B+B\Theta^{-2}B)+\tr (C\Theta^2 C^T+2C\Sigma C+C^T\Theta^{-2}C),$$
where
$\Sigma=(\Theta^2+\Theta^{-2}-\Omega\Phi-\Phi\Omega^{-1})/2=\Theta^2-\Omega\Phi$
(the proof of the second ``=" is nontrivial but will be omitted
here).

For illustration, we carry out the Hessian analysis for critical
submanifolds that contain only type I and II singular values,
where the corresponding $\Theta$ and $\Sigma$ are diagonal.
Moreover, we assume that $\om_i>1$ for all $i=1,\cdots,N$, and the
spectrum of $E_d$ is so widely spaced that $\om_i\nprec \om_j$ for
any $\om_i<\om_j$ (the other cases not involving type III singular
values can be dealt with as well but are relatively cumbersome).
Now suppose that the diagonal elements in $\Theta$ are ordered as
$$diag\{\om_1 I_{m_1'},\om_1^{1/3} I_{m_1''};\cdots,;\om_s I_{m_s'},\om_s^{1/3} I_{m_s''}\}.$$
Then the Hessian form can be decomposed into
$\h(A,B,C)=\h_1(A,B,C)+\h_2(A,B,C)$ with
\begin{equation}
\begin{array}{l}
\h_1(A,B,C)= \sum_{j=1}^{N}\Big[(d_j^{-2}+2\sigma_j+d_j^{2})c_{jj}^2+(d_ja_{jj}-\sigma_jd_j^{-1}b_{jj})^2\Big]\\
  +\sum_{1\leq i<j\leq
   {N}}\left\{\left[(d_i^{-2}+d_j^2)^{\frac{1}{2}}c_{ij}+\frac{(\sigma_i+\sigma_j)c_{ji}}{(d_i^{-2}+d_j^2)^{\frac{1}{2}}}\right]^2+
\left[(d_i^{2}+d_j^2)^{\frac{1}{2}}a_{ij}+\frac{(\sigma_i+\sigma_j)b_{ij}}{(d_i^{2}+d_j^2)^{\frac{1}{2}}}\right]^2\right\},\\
 \end{array}\end{equation}
\begin{equation}
\begin{array}{l}
\h_2(A,B,C)=
\sum_{j=1}^N(1-\sigma_j^{2})d_j^{-2}b_{jj}^2\\
+\sum_{1\leq i<j\leq
   N}\left\{
\left[(d_i^2+d_j^{-2})-\frac{(\sigma_i+\sigma_j)^2}{d_i^{-2}+d_j^2}\right]c_{ji}^2+\left[(d_i^{-2}+d_j^{-2})-\frac{(\sigma_i+\sigma_j)^2}{d_i^{2}+d_j^2}\right]b_{ij}^2\right\}
, \end{array}\end{equation} where $a_{ij}$, $b_{ij}$ and $c_{ij}$
are matrix elements of $A$, $B$ and $C$; $d_j$ and $\sigma_j$ are
diagonal matrix elements of $\Theta$ and $\Sigma$. The expressions
consisting of square terms of independent variables actually
represent the local coordinate system in which the HQF is
diagonalized. This can be used to count the number of positive
(negative or zero) Hessian eigenvalues by examining the signs of
these square terms.

The first part $\h_1(X)$ contains $N^2+N$ positive definite terms
with respect to any choice of $X$, and hence it provides $N^2+N$
positive Hessian eigenvalues. The $N^2$ terms in the (positive
indefinite) second part $\h_2(X)$ needs to be further analyzed. It
is easy to see that the coefficients of the first $N$ terms
$(1-\sigma_j^{2})d_j^{-2}b_{jj}^2$, $j=1,\cdots,N$, are positive
for type I singular values $d_j$ where $\sigma_j=0$, and negative
for type II singular values $d_j$ where
$\sigma_j=d_j^2+d_j^{-2}\geq2$. They provide $N'=\sum_{i=1}^s
m_i'$ positive and $N''=\sum_{i=1}^s m_i''$ Hessian eigenvalues.

The signs for the remaining terms are determined by the
discriminants
$\Delta'_{ij}=(d_i^2+d_j^{-2})({d_i^{-2}+d_j^2})-{(\sigma_i+\sigma_j)^2}$
and
$\Delta''_{ij}=(d_i^2+d_j^2)({d_i^{-2}+d_j^{-2}})-{(\sigma_i+\sigma_j)^2}$,
where $i<j$, whose signs correspond to that of Hessian eigenvalues
in the coordinates of $c_{ij}$ and $b_{ij}$, respectively.

(1) When both $d_i$ and $d_j$ are of type I, $\sigma_i=\sigma_j=0$
and hence both $\Delta'_{ij}$ and $\Delta''_{ij}$ are positive.
This produces $N'(N'-1)$ positive Hessian eigenvalues.

(2) When both $d_i$ and $d_j$ are of type II, the value of
$\Delta'_{ij}$ is:
\begin{eqnarray*}
  \Delta'_{ij} &=& (d_i^2+d_j^{-2})({d_i^{-2}+d_j^2})-{(d_i^2+d_i^{-2}+d_j^2+d_j^{-2})^2} \\
   &=&
   -[(d_i^{-2}+d_j^2)^2+(d_i^2+d_j^{-2})({d_i^{-2}+d_j^2})+{(d_i^2+d_j^{-2})^2}]<0,
\end{eqnarray*}
from which we can see that the corresponding Hessian eigenvalues
are all negative. The same holds for $\Delta''_{ij}$.  In total,
this produces $N''(N''-1)$ negative Hessian eigenvalues.

(3) When $d_i$ is of type I and $d_j$ is of type II, the
discriminants become:
\begin{eqnarray*}
\Delta'_{ij}&=&(d_i^2+d_j^{-2})({d_i^{-2}+d_j^2})-{(d_j^2+d_j^{-2})^2}=d_j^{-4}(d_i^2-d_j^2)(d_j^{6}-d_i^{-2}),\\
\Delta''_{ij}&=&(d_i^2+d_j^{2})({d_i^{-2}+d_j^{-2}})-{(d_j^2+d_j^{-2})^2}=d_j^{4}(d_i^2-d_j^{-2})(d_j^{-6}-d_i^{-2}).
\end{eqnarray*}
Because the $d_i$ and $d_j$ are always chosen to be greater than
$1$, it is easy to see that $d_j^{6}-d_i^{-2}>0$ and
$d_i^2-d_j^{-2}>0$ except when $d_i=d_j=1$. Hence the signs of the
discriminants are determined by
$$\Delta'_{ij}\sim d_i^2-d_j^2=\om_i^2-\om_j^{2/3},
~~~~\Delta''_{ij}\sim d_j^{-6}-d_i^{-2}=\om_j^{-2}-\om_i^{-2}.
$$ $\Delta'_{ij}$ is positive only when $\om_i\prec\om_j$, and otherwise negative. However, by assumption this holds only when
$\om_i=\om_j$, which brings $\sum_{\alpha}m'_\alpha m''_\alpha$
positive Hessian eigenvalues and $\sum_{\alpha< \beta}m'_\alpha
m''_\beta$ negative Hessian eigenvalues. The discriminant
$\Delta'_{ij}\leq 0$ for $\om_i\leq\om_j$, which brings
$\sum_{\alpha< \beta}m'_\alpha m''_\beta$ negative, and
$\sum_{\alpha}m'_\alpha m''_\alpha$ zero Hessian eigenvalues.

(4) When $d_i$ is of type II and $d_j$ is of type I, we may
derive:
\begin{eqnarray*}
\Delta'_{ij}&=&(d_i^2+d_j^{-2})({d_i^{-2}+d_j^2})-{(d_i^2+d_i^{-2})^2}=d_i^{-4}(d_j^2-d_i^2)(d_i^{6}-d_j^{-2}),\\
\Delta''_{ij}&=&(d_i^2+d_j^{2})({d_i^{-2}+d_j^{-2}})-{(d_i^2+d_i^{-2})^2}=d_i^{4}(d_j^2-d_i^{-2})(d_i^{-6}-d_j^{-2}),
\end{eqnarray*}which lead to the similar criteria
$$\Delta'_{ij}\sim d_j^2-d_i^2=\om_j^2-\om_i^{2/3}>0,
~~~~\Delta''_{ij}\sim d_i^{-6}-d_j^{-2}=\om_i^{-2}-\om_j^{-2}>0.
$$The Hessian eigenvalues in this case are all positive and its number is $2\sum_{\alpha< \beta}m''_\alpha m'_\beta$.

In conclusion, the total number of positive, negative and null
Hessian eigenvalues can be summated as follows:
\begin{eqnarray}
  \D_0 &=& \sum_{\alpha=1}^r
  m'_\alpha m''_\alpha, \label{d0}\\
  \D_+ &=& N^2+N+N'^2+\sum_{\alpha=1}^r
  m'_\alpha m''_\alpha+2\sum_{\alpha< \beta }m'_\alpha m''_\beta, \label{d+}\\
\D_- &=& N''^2+2\sum_{\alpha<\beta}m''_\alpha m'_\beta.\label{d-}
\end{eqnarray}
From these formulas, it is easy to see that, among these critical
submanifolds, there is only one local minimum $S^*=W$ in the
landscape whose singular values are all of type I. Thehe rest of
them are all saddle submanifolds because both $\D_+$ and $\D_-$
are nonzero. The same conclusion can be drawn for other critical
submanifolds that have no type III blocks.

The Hessian analysis for critical submanifolds involving type III
singular values is more complicated and any analytic formula is
not available so far. However, it is not difficult to prove that
they are all saddle submanifolds. So we may conclude the main
theorem in this paper:
\begin{theorem}
The optimization problem (\ref{symplectic landscape}) has a unique
minimum $S^*=W$ and the rest of the critical submanifolds are all
saddles.
\end{theorem}
{\bf Proof:} It is sufficient to prove that critical submanifolds
involving type III singular values have saddle structures, i.e.,
the corresponding Hessian form is neither positive or negative
definite, or equivalently, there exist some $X'\neq 0$ and
$X''\neq 0$ such that $\h(X')>0$ and $\h(X'')<0$. Let
$\mathcal{M}$ be such a critical submanifold, and
$\Sigma=diag\{\Sigma_0,\Sigma_1\}$, where $\Sigma_0$ is a
$2k_i\times 2k_i$ type III block and $\Sigma_2$ contains the rest
blocks. Choose a particular $A=diag\{A_0;0\}$ (similarly for $B$
and $C$), where $A_0$ corresponds to $\Sigma_0$ and $0$ to
$\Sigma_1$, such that the resulting Hessian quadratic form are
irrelevant to $\Sigma_1$, i.e.,
$$\h(A_0,B_0,C_0)=\tr(A_0\Theta_0^2A_0-2A_0\Sigma_0 B_0+B_0\Theta_0^{-2}B_0)+\tr (C_0\Theta_0^2 C_0^T+2C_0\Sigma_0 C_0
+C_0^T\Theta_0^{-2}C_0).$$ Here the
sub-block $$\Sigma_0=\left(%
\begin{array}{cc}
  (d_i^2-\gamma_i\cos x_i)I_{k_i} & -\gamma_i \sin x_iI_{k_i} \\
  -\gamma^{-1}_i \sin x_iI_{k_i} & (d_i^2+\gamma_i^{-1}\cos x_i)I_{k_i} \\
\end{array}%
\right)$$ where $\gamma_i=(\om_\alpha\om_\beta)^{\frac{1}{2}}$
with $\om_{\alpha}$ and $\om_\beta$ being the pair of singular
values of $W$ that generates the singular value
$d_i=(\om_\beta/\om_\alpha)^{1/2}$. The matrix
$\Theta_0=d_iI_{2k_i}$.

Now choose $A_0=I_2$, $B_0=\lmd I_2$ and $C_0=0$, where $\lmd$ is
to be determined. Then
$$\h(A_0,B_0,C_0)=\tr(\Theta_0^2+2\lmd\Sigma_0
+\lmd^2\Theta_0^{-2})=2k_i[(d_i^2+\lmd
d_i^{-2})-\lmd(2d_i^2-(\gamma_i-\gamma_i^{-1})\cos x_i)].$$
According to the definition $\cos
x_i=(d^2_i-d^{-2}_i)/(\gamma_i-\gamma^{-1}_i)$, the Hessian can be
simplified as $\h(A_0',B_0',C_0')=2k_i(1-\lmd)d_i^2$. So the
corresponding $\h(X_0)$ is positive (resp., negative) when
$\lmd<1$ (resp., $\lmd>1$ ), which implies that the Hessian is
neither positive nor negative definite. End of proof.

\section{Critical Landscape Topology constrained on the compact
symplectic group}

Carrying out optimal control field searches over only the compact
subgroup O$\Sp(2N,\R)$ is also important in many circumstances,
e.g., using only linear quantum optics to search for a symplectic
quantum gate \cite{WuRaj2007}. The derivation of the topology is
similar to that for the landscapes on $\U(N)$ described above,
since $\OSp(2N,\R)$ is isomorphic to $\U(N)$ \cite{ArvMuk1995}.
The Lie algebra of $\OSp(2N,\R)$ consists of matrices of the form
$$o\spl(2N,\R)=\{A=JY~\Big|~Y^T=Y,~JY=YJ\}.$$
The condition for $S$ to be a critical point in the constrained
landscape is
$$\tr(W^TSJY)=0,\quad \forall~ JY\in o\spl(2N,\R),$$
which can only hold if the matrix $W^TSJ$ is an element of the
space complementary to that of $B$, which is equivalent to
requiring that $W^TS$ is an element of the Jordan algebra of
O$\Sp(2N,\R)$). $W^TSJ$ must then simultaneously satisfy the two
conditions
$$W^TSJ=-(W^TSJ)^T=JS^TW, \quad J(W^TSJ)=(W^TSJ)J,$$
which implies that $W^TS=S^TW$. This equation can be rearranged to
give $S=W\sqrt{I_{2N}}$, where $\sqrt{I_{2N}}$ must lie within the
group $\OSp(2N,\R)$. Because the cost functional is invariant with
respect to the conjugation action of O$\Sp(2N,\R)$, the solutions
correspond to a set of O$\Sp(2N,\R)$ orbits, i.e., $S^*=WR^TD_mR$,
where $R\in\OSp(2N,\R)$ and
$$D_m=\left(%
\begin{array}{cccc}
  -I_m & & &  \\
   &  I_{N-m} & & \\
   &   &  -I_m &\\
   &   &   &  I_{N-m} \\
\end{array}%
\right),\quad 0\leq m\leq N.
$$
There are then $N+1$ solutions, with values of the cost functional
$\J=0,8,16,...,8N$. The minimum and maximum values of $\J$
correspond to $S=W$ and $S=-W$, respectively. The critical
manifolds can be expressed as Grassmannian cosets
$\mathcal{M}=\OSp(2N,\R)/$\G$(D_m)$, where $\G(D_m) =
\OSp(2m,\R)\times \OSp(2(N-m),\R)$, so that
$$G(m,N)=\frac{\OSp(2N,\R)}{\OSp(2m,\R)\times \OSp(2(N-m),\R)}.$$

The HQF can be calculated by parameterizing the argument of
$\J(S)=\tr(W^TS)$ via $Se^{JY}$ as in the above. Taylor expanding
the landscape function and keeping only the second-order term, we
get the HQF,
$$\h(Y)=\tr[W^TS^*(JY)^2]=\tr[R^TD_mR(JY)^2]=\tr[(JRYR^T)^TD_m(JRYR^T)].$$ Let $X=RYR^T$, which still satisfies the
conditions $X^T=X$ and $XJ=JX$. $X$  can be expressed in the form
$$X=\left(%
\begin{array}{cc}
  A&   C \\
 -C&   A \\
\end{array}%
\right)
$$
where $A^T=A$ and $C^T=-C$ are $N$-dimensional matrices. Let
$a_{ij} = a_{ji}$ and $c_{ij} = - c_{ji}$ are the matrix elements
of $A$ and $C$. Since any $S^*$ is represented by a corresponding
matrix $D$, we obtain the following polynomial expression for the
HQF:
$$H(X) =
-2\sum_{j=1,N}a_{jj}^2\delta_j-2\sum_{1\leq i<j\leq
N}(a_{kl}^2+c_{kl}^2)(\delta_k+\delta_l),$$where $\delta_j$ is the
$j$-th diagonal element of $D_m$. It can then be verified that the
landscape on the homogenous compact symplectic group has identical
critical topology to the (unitary) transformation landscape on
$\U(N)$ \cite{Mike2006b}, with the following breakdown of Hessian
eigenvalues for $m=0,\cdots,N$:
\begin{equation}
\D_+ = (N-m)^2,\quad \D_- = m^2,\quad \D_0= 2m(N-m).
\end{equation}

As in the case of the optimization over the full symplectic group,
the critical topology for compact target symplectic gates also
consists of orbits of orthogonal symplectic groups, whose numbers
of Hessian eigenvalues are
\begin{equation}
\D_+ =N^2+N+ (N-m)^2,\quad \D_- = m^2,\quad \D_0= 2m(N-m),
\end{equation}where $m$ is defined in the standardized block $P_0$
in Section III. This shows that the critical topologies are very
close between the full symplectic group and its compact subgroup,
except for the $N^2+N$ difference in the number of positive
Hessian eigenvalues. By this difference, the non-optimal critical
points for $\Sp(2N,\R)$ are all saddle points, while one of
$\OSp(2N,\R)$ is a minimal point.

\section{Examples}
Consider the SUM gate that has one two-fold degenerate singular
value $\om=(\sqrt{5}+1)/2$. The analysis predicts that there are 4
critical submanifolds for this gate. The first one is the global
minimum point $S^*_1={\rm SUM}$, whose characteristic matrix
$P_1=diag\{\om,\om^{-1} , \om^{-1},\om\}$ contains one type I
block ($m'=2$, $m''=0$).

The second critical submanifold is an isolated saddle point, whose
characteristic matrix
$P_2=diag\{-\om^{-1/3},-\om^{1/3},-\om^{1/3},-\om^{-1/3}\}$
contains one type II block ($m'=0$, $m''=2$), and the
corresponding critical point is
$$S^*_2=UP_2V=\left(%
\begin{array}{cccc}
  -0.906  &  0.614    &     0   &      0 \\
     -0.292 &  -0.906  &       0 &        0 \\
           0  &       0  & -0.906  & 0.292 \\
           0   &      0  & -0.614  & -0.906 \\
\end{array}%
\right).$$

As to the third one, where $P_3=diag\{\om,-\om^{1/3},
\om^{-1},-\om^{-1/3}\}$ contains one type I and one type II blocks
($m'=1$, $m''=1$), the corresponding critical submanifold is
one-dimensional as the orbit of the $O(2)$ symmetry group of $E$.
Parameterize $\G(E_d)$ as:
$$R(\theta)=\left(\begin{array}{cccc}
  \cos\theta  &  \sin\theta    &     0   &      0 \\
  \mp\sin\theta &  \pm\cos\theta  &       0 &        0 \\
           0  &       0  & \cos\theta  & \mp\sin\theta \\
           0   &      0  & \sin\theta  & \pm\cos\theta \\
\end{array}%
\right), ~~~~\theta\in[0,2\pi),$$ where the signs $\pm$ correspond
to the two disjoint parts of $O(2)$. The critical submanifold can
be expressed as an orbit of $O(2)$ group:
$$S^*_{3}={\small \left(%
\begin{array}{cccc}
0.152\cos\theta+0.047 &    0.990\cos\theta+0.307 &
0.307\sin\theta&          0.953\sin\theta \\
1.141\cos\theta+0.354& 0.152\cos\theta+0.047& -0.953\sin\theta&
0.646\sin\theta \\
 -0.646\sin\theta&         -0.953\sin\theta&
0.047-0.152\cos\theta&   -0.354+1.141\cos\theta \\
0.953\sin\theta&         -0.307\sin\theta& -0.307+0.990\cos\theta&
0.047-0.152\cos\theta\\
\end{array}%
\right),}$$where the orbits of the two disjoint parts coincide with
each other.

The last critical submanifold contains a type III block ($m'=0$,
$m''=0$ and $m'''=1$). This block and its corresponding critical
matrix are given by:

$$P_4=\left(%
\begin{array}{cccc}
 0  & 1 & 0 & 0 \\
 1  & 0 & 0 & 0 \\
 0  & 0 & 0 & 1 \\
 0  & 0 & 1 & 0 \\
\end{array}%
\right),\quad    S^*_4=\left(%
\begin{array}{cccc}
    1  &  0 &  0  &  0\\
    0  &  -1 &  0  &  0\\
    0  &  0 &  1  & 0\\
    0  &  0 &  0  & -1 \\
\end{array}%
\right).$$

The Hessian analysis can be done for the first three critical
points (submanifolds) with formula given in Section III. Here we
exemplify the Hessian analysis with the critical point $S^*_4$,
for which we don't have an explicit counting formula yet. Using
expression (\ref{HQF-X}) and the decomposition of $X$ into
$(A,B,C)$, we have $\h(X)=\h_1(A,B)+\h_2(C)$, where
\begin{eqnarray*}
  \h_1(A,B) &=& a_{11}^2+2a_{12}^2+a_{22}^2+b_{11}^2+2b_{12}^2+b_{22}^2
  -2a_{11}b_{11}-4a_{12}b_{12}+2\om (a_{11}b_{12}+a_{12}b_{22})\\
&&  +2\om^{-1}(a_{12}b_{11}+a_{22}b_{12}) -2a_{22}b_{22}\\
\h_2(C)&=&4c_{11}^2+2c_{12}^2+2c_{21}^2+4c_{22}^2 -2\om
(c_{11}c_{21} + c_{22}c_{21})-2\om^{-1}(c_{12}c_{22}+
c_{11}c_{12})+4c_{21}c_{12}
\end{eqnarray*}

Denoting by ${\bf\rm
x}=(a_{11},a_{12},a_{22},b_{11},b_{12},b_{22},c_{11},c_{12},c_{21},c_{22})$
the vector of independent variables, the Hessian form can be
expressed as a quadratic polynomial ${\bf\rm x}^T \mathcal{Q}
{\bf\rm x}$, where
$$\mathcal{Q}=\left(%
\begin{array}{cccccccccc}
  1 & 0 & 0 & -1 & \om & 0 & 0 & 0 & 0 & 0 \\
  0 & 2 & 0 & \om^{-1} & -2 & \om & 0 & 0 & 0 & 0 \\
  0 & 0 & 1 & 0 & \om^{-1} & -1 & 0 & 0 & 0 & 0 \\
  -1 & \om^{-1} & 0 & 1 & 0 & 0 & 0 & 0 & 0 & 0 \\
  \om & -2 & \om^{-1} & 0 & 2 & 0 & 0 & 0 & 0 & 0 \\
  0 & \om & -1 & 0 & 0 & 1 & 0 & 0 & 0 & 0 \\
  0 & 0 & 0 & 0 & 0 & 0 & 4 & -\om^{-1} & -\om & 0 \\
  0 & 0 & 0 & 0 & 0 & 0 & -\om^{-1} & 2 & 2 & -\om^{-1} \\
  0 & 0 & 0 & 0 & 0 & 0 & -\om & 2 & 2 & -\om \\
  0 & 0 & 0 & 0 & 0 & 0 & 0 & -\om^{-1} & -\om & 4 \\
\end{array}%
\right)$$ is a block-diagonal 10-dimensional symmetric matrix.
Numerical calculation shows that upper block corresponding to
$\h_1(A,B)$ offers 4 positive and 2 negative Hessian eigenvalues;
the lower block corresponding to $\h_2(C)$ offers 2 positive and 1
negative Hessian eigenvalues. Hence the HQF has 7 positive and 3
negative eigenvalues.

In summary, there are a total of six critical submanifolds
including 3 isolated points and two one-dimensional manifolds. The
Hessian analyses are summarized in Table I.
\begin{center}
\begin{table}[h]
\begin{tabular}{c|c|c|c|c|c}
\hline
No. & Critical value  & ~~$\D_0$~~ & ~~$\D_+$~~ & ~~$\D_-$~~ & ~type~ \\
\hline
1&        0  &   0 &  10 &        0 & minimum \\
2&   18.623  &   0 &  6 &   4 & saddle \\
3&  9.311    &   1 &  8 &   1 & saddle \\
4&  10       &   0 &  7 &   3 & saddle \\
\hline                                   
\end{tabular}
\caption{Landscape characteristics for the SUM gate.}
\end{table}
\end{center}

\section{Conclusion}
We have resolved the critical solutions for least square problems
on the symplectic group. The critical topology of this nonlinear
optimization problem over a noncompact Lie group was shown to be
of high complexity compared to that of analogous problems on
compact Lie groups. However, the topology is still devoid of
multiple local extrema, and the critical solutions consist of a
finite number of critical submanifolds which are within a bounded
region. These results have important applications to the study of
control landscapes \cite{Raj2007} for classical mechanical systems
or continuous variable quantum computation systems, implying that
the search of optimal controls would encounter no essential
obstructions.

Due to the noncompactness of the symplectic group, the optimal
implementation of symplectic transformations (or symplectic gates
in continuous variable quantum computation) might be more
inefficient than that of unitary transformations (e.g., those
applied in discrete variable quantum computation). Nonetheless,
recent OCT simulations using this objective function
\cite{WuRaj2007} verify the prediction that local gradient-based
algorithms will converge due to the lack of local traps in the
landscape.

\appendix\section{Stabilizers of symplectic matrices}
Here we give the structures of stabilizers of several kinds of
symplectic matrices encountered in this paper. The blocks
$D_a=diag\{aI_n;a^{-1}I_n\}$ will be frequently encountered
corresponding to a reciprocal pair of singular values $a$ and
$a^{-1}$. We can substitute the standard form (\ref{U-S map}) of
$R$ into the definition, which gives
$$\left(%
\begin{array}{cc}
  aX & aY \\
  -a^{-1}Y & a^{-1}X \\
\end{array}%
\right)=\left(%
\begin{array}{cc}
  aX & a^{-1}Y \\
  -aY & a^{-1}X \\
\end{array}%
\right).$$ It is easy to see that, when $a=1$, any matrix
$R\in\OSp(2n,\R)$ is in the stabilizer. For $a\neq 1$, $Y$ has to
be zero, and hence leaves $R=diag\{X;X\}$ where $X\in O(n)$. So,
the stabilizer for such $D$ is isomorphic to $O(n)$. 



\end{document}